\documentclass[useAMS,usenatbib]{mn2e}

\usepackage{psfig}
\usepackage{graphicx}
\usepackage[english]{babel}  
\usepackage{txfonts}
\usepackage{url}

\newcommand\apj{ApJ}

\newcommand\aap{A\&A}
\newcommand\apjl{ApJ}
\newcommand\nat{Nature}
\newcommand\apjs{ApJS}
\newcommand\mnras{MNRAS}
\newcommand\pasj{PASJ}
\newcommand\aaps{A\&AS}

\title[Modeling of the Vela shrapnel]{Hydrodynamic modelling of ejecta shrapnel in the Vela supernova remnant}
\author[M. Miceli et al.]{M. Miceli$^{1,2}$\thanks{E-mail:
miceli@astropa.unipa.it)}, S. Orlando$^{2}$, F. Reale$^{1,2}$, F. Bocchino$^{2}$, G. Peres$^{1,2}$\\
$^{1}$Dipartimento di Fisica, Universit\`a di Palermo, Piazza del Parlamento 1, 90134 Palermo, Italy\\
$^{2}$INAF - Osservatorio Astronomico di Palermo, Piazza del Parlamento 1, 90134 Palermo, Italy }
\begin{document}

\date{Accepted . Received ; in original form }

\pagerange{\pageref{}--\pageref{}} \pubyear{2002}

\maketitle

\label{firstpage}

\begin{abstract}
Many supernova remnants (SNRs) are characterized by a knotty ejecta structure. The Vela SNR is an excellent example of remnant in which detached clumps of ejecta are visible as X-ray emitting bullets that have been observed and studied in great detail.
We aim at modelling the evolution of ejecta shrapnel in the Vela SNR, investigating the role of their initial parameters (position and density) and addressing the effects of thermal conduction and radiative losses.
We performed a set of 2-D hydrodynamic simulations describing the evolution of a density inhomogeneity in the ejecta profile. We explored different initial setups.
We found that the final position of the shrapnel is very sensitive to its initial position within the ejecta, while the dependence on the initial density contrast is weaker. Our model also shows that moderately overdense knots can reproduce the detached features observed in the Vela SNR. Efficient thermal conduction produces detectable effects by determining an efficient mixing of the ejecta knot with the surrounding medium and shaping a characteristic elongated morphology in the clump.
\end{abstract}


\section{Introduction}
The ejecta in supernova remnants (SNRs) drive the exchange of mass and the chemical evolution of the galactic medium. The structure of SNR ejecta has been proved to be knotty, and several clumps have been observed at different wavelegths in remnants of core-collapse supernovae, as G292.0$+$1.8 (\citealt{phs04}), Puppis A (\citealt{kmt08}), and Cas A, where knots have been observed also beyond the main shock front (\citealt{fhm06}, \citealt{hf08}, \citealt{drs10}).

The Vela SNR, being the nearest SNR, represents a privileged target for this kind of studies, since it is possible to observe fine structures down to small physical scales. Despite the bulk of the X-ray emission of the Vela SNR is associated with the post-shock interstellar medium, X-ray emitting ejecta have also been observed. In particular, six protruding features, with characteristic boomerang morphology, (labelled Shrapnel A-F) have been identified in the Vela SNR by \citet{aet95}, who argued an ejecta origin for these structures which appear to be detached from the remnant. The association with ejecta fragments has been supported by more recent observations performed with $Chandra$, \emph{XMM-Newton}, and $Suzaku$. 
The analysis of the \emph{XMM-Newton} observation of Shrapnel D, has revealed that O, Ne, and Mg abundances are significantly larger than solar (\citealt{kt05}). A similar abundance pattern has been observed with $Suzaku$ in Shrapnel B \citep{yk09}, but in this case the overabundances of the lighter elements are less prominent, suggesting more effective mixing with the interstellar medium (ISM).
A $Chandra$ observation of Shrapnel A, whose projected distance from the center of the remnant is larger by $\sim 20\%$ than Shrapnel D, reveals instead oversolar Si$:$O ratios (\citealt{mta01}). Significant Si overabundance (Si$\sim3$) has been confirmed by \citet{kt06}, who analyzed an \emph{XMM-Newton} observation, finding solar or subsolar values for the O, Ne, Mg, and Fe abundances.
These results show differences in the chemical composition between Shrapnel A and Shrapnel B and D.
In the northern rim of the Vela shell, \citet{mbr08} discovered new X-ray emitting clumps of ejecta whose projected position is behind the main shock front. The relative abundances (O:Ne:Mg:Fe) of these new shrapnel are in good agreement with those observed in Shrapnel D. Similar abundance pattern have been observed also by \citet{lsd08}, who found ejecta-rich plasma in the direction of the Vela X pulsar wind-nebula.

The present day morphology of SNRs and the structure of ejecta are believed to reflect the physical characteristics of the SN explosion (e.~g., intrinsic asymmetries of the explosion, interaction of the early blast with the inhomogeneities of the circumstellar medium, physical processes in the aftermath of the explosion, etc.) and their detailed study promises
to contribute to our understanding of the SN explosion physics. In the light of these considerations, it is then interesting to model the evolution of the ejecta knots to understand how the current position and chemical properties of the shrapnel in the Vela SNR depend on the physical conditions at the supernova explosion and on the dynamics of the explosion itself.

The evolution of dense, supersonic clumps of SN ejecta running in a uniform medium has been studied by \citet{ajr94} and \citet{jkt94}, who identified three main stages of evolution: a bow-shock phase, an instability phase and a dispersal phase. However, these models do not describe in detail the interaction of the clump with the remnant (post-shock medium, main shock, reverse shock) and do not include important physical effects (as thermal conduction and radiative cooling). \citet{cpr96} included radiative losses in their 2-D models, but they focussed on the interaction of a knot with a very small supernova remnant ($\sim 10^{17}$ cm, i. e. more than 100 times smaller than the Vela SNR) evolving in an extremely dense medium ($10^7$ cm$^{-3}$). A hydrodynamic model (without thermal conduction and radiative cooling) specifically tuned for the Vela SNR has been developed by \citet{wc02} (hereafter WC02) who followed the evolution of a shrapnel by using 2-D simulations in spherical coordinates (because of the 
geometry of their simulations, the shrapnel are modeled as toroidal structures with very large masses). WC02 did not model the early evolution of the ejecta knot, but started their simulations at the time $t_{rev}$, corresponding to the first interaction of the shrapnel with the reverse shock front. They explored different values of $t_{rev}$ and of the density contrast between the shrapnel and the surrounding ejecta, $\chi$, and found that, in order to produce an observable protrusion on the shock front (like that observed in Shrapnel A-F), a very high density contrast ($\chi\sim 1000$) is necessary. With lower density contrasts ($\chi<100$), the shrapnel are rapidly decelerated and fragmented by hydrodynamic instabilities and the observed features cannot be reproduced (for the effects of hydrodynamic instabilities on shocked clouds see also \citealt{kmc94} and \citealt{opr05}). Large density inhomogeneities in the clumps are difficult to explain in a core-collapse SN explosion. WC02 argued 
that a model that includes the effects of radiative cooling may show that lower values of $\chi$ are needed to match the 
observed protrusions.
Also, WC02 do not include in their model the effects of thermal conduction that, as shown by \citet{opr05}, can efficiently suppress the hydrodynamic instabilities, thus allowing the shrapnel to overcome the main shock-front without being disrupted. 
Recently, the evolution of knotty ejecta in a Type Ia SNR has been modelled by \citet{obm12} (hereafter O12), who found that small clumps with initial $\chi<5$ can reach the SNR shock front after $\sim1000$ yr. Nevertheless, these ejecta knots are then rapidly eroded and do not produce significant protrusions in the SNR shock front, thus being unable to reproduce the features observed in the Vela SNR.   
 
Here we present a set of 2-D hydrodynamic simulations of the evolution of an (initially spherical) ejecta shrapnel in the Vela SNR. We include in our model both thermal conduction and radiative cooling and explore different values of $\chi$ and of the initial position of the shrapnel in the ejecta profile. We aim at addressing the role of thermal conduction and radiative cooling and at understanding how the initial properties of the shrapnel influence its evolution. We also aim at evaluating whether values of $\chi$ lower than 100 ($\chi=10-50$) can reproduce the observed features.

The paper is organized as follows: the hydrodynamic model is described in Sect. \ref{Hydrodynamic modeling}, the results of the simulations are shown in Sect. \ref{Results}, and our conclusions are discussed in Sect. \ref{Conclusions}.

\section{Hydrodynamic modeling}
\label{Hydrodynamic modeling}

\subsection{Initial conditions and model equations}
\label{initcond}

We model the evolution of a shrapnel in a SNR by performing a set of 2-D simulations in a cylindrical coordinate system $(r, ~ z)$, assuming axial symmetry. The system setup consists of a spherically symmetric distribution of ejecta with initial kinetic energy $K=10^{51}$ erg (the initial thermal energy is only the $0.2\%$ of $K$) and mass $M_{ej}=12$ M$_{\sun}$ (representing the initial blast wave), where we place a dense, spherical, knot (the shrapnel) in pressure equilibrium with the surrounding ejecta and with central coordinates $(0,~R_s)$. 

The radial density profile of the ejecta consists of two power-law segments ($\rho\propto r^{-m}$ on the inside and $\rho\propto r^{-b}$ on the outside), in agreement with the density structure in a core-collapse SN described by \citet{che05}. For our simulations, we use $m=1$ and $b=11.2$, and the position of the transition between the flat and steep regimes is derived by following \citet{che05}. The initial velocity of the ejecta increases linearly (up to $6\times10^{8}$ cm/s) with their distance from the center. The maximum velocity is reached at the initial radius of the ejecta, i. e., $R^{0}_{ej}=4.5\times10^{18}$ cm. These values correspond to $\sim240$ yr after the explosion, appropriate for the relatively late stages of the SNR evolution that we address in this study. In fact, the starting time of our simulations corresponds to only $\sim2\%$ of the Vela SNR age and the shrapnel reaches the reverse shock $\sim2500-3000$ yr after the explosion in all our simulations (i. e., the system has enough time 
to evolve, before the interaction of the shrapnel with the SNR reverse shock occurs). 
We then conclude that our simulations can provide a realistic description of the actual conditions in the Vela SNR. 

\begin{figure}
  \centerline{\hbox{	 
      \psfig{figure=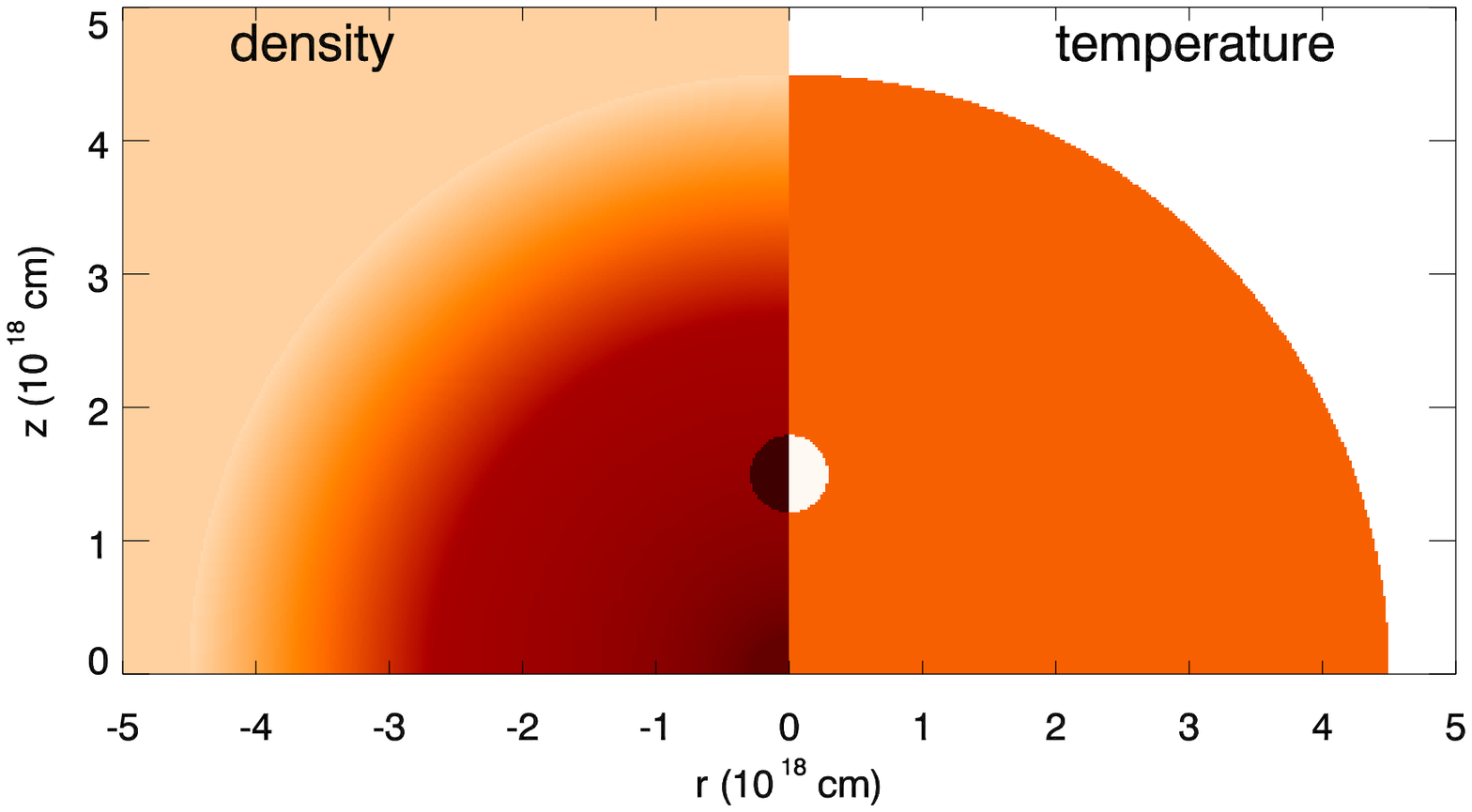,width=7cm}}}
 \centerline{}      
 \centerline{}      
 \centerline{}      
  \centerline{\hbox{	 
      \psfig{figure=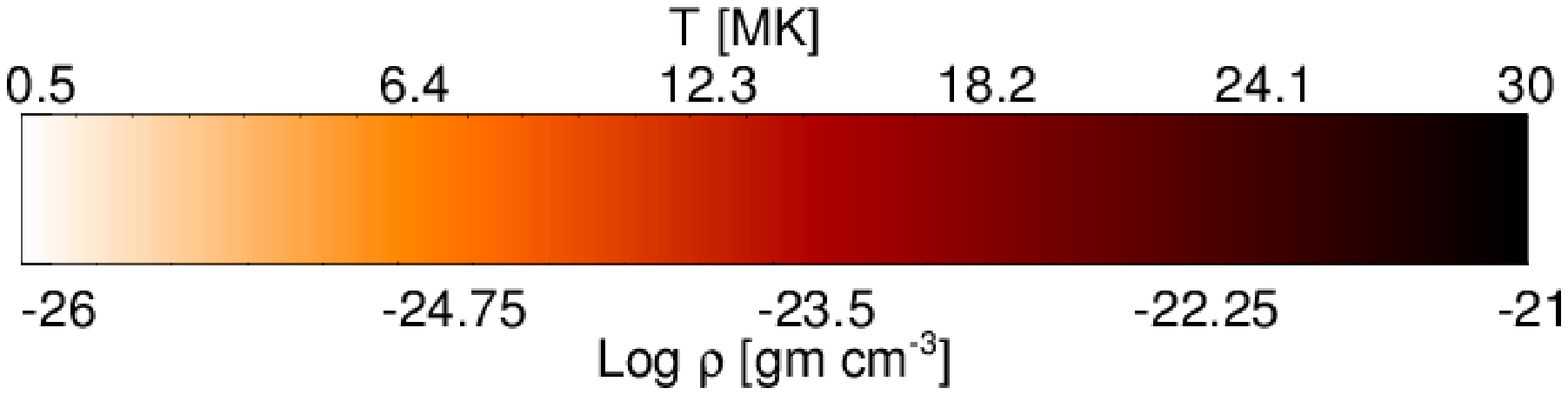,width=5cm}}}
 \caption{Density (\emph{left panel}) and temperature (\emph{right panel}) 2-D cross-sections through the $(r,~z)$ plane showing an example of the initial conditions of our simulations (namely, $R1/3-CHI20$, see Table \ref{tab:setups}). The system consists of an expanding spherically symmetric distribution of ejecta where we place a dense, isobaric, and spherical knot. In this case the knot is 20 times denser than the surrounding ejecta.}
\label{fig:setup}
\end{figure}

The initial mass of the shrapnel is $0.05 ~ M_{ej}$ ($1.19\times10^{33}$ g), its density is $\chi$ times larger than that of the surrounding ejecta at distance $R_s$ from the center\footnote{The size of the clump varies accordingly.}, and its velocity is the same as that of the ejecta at distance $R_s$ from the center. We aim at showing that the detached shrapnel observed in the Vela SNR can be the result of moderately overdense clumps of ejecta originating in relatively internal layers. We explored different values of $\chi$ and of $R_s$, namely $\chi=10,~20,~50$, and $R_s=1/6~R^{0}_{ej},~1/3~R^{0}_{ej}$. Figure \ref{fig:setup} shows the initial density and temperature conditions for the case with $\chi=20$ and $R_s=1/3~R^{0}_{ej}$. The simulation setups discussed in this paper are summarized in Table \ref{tab:setups}.

Vela SNR is the result of a core-collapse SN and we expect the ambient medium to be ``perturbed'' by the wind residuals of the massive progenitor star. However, we assume for simplicity a uniform ambient medium as in WC02, since here we are not interested in modeling the details of the remnant evolution. The final (i.~e. after 11000 yr, the age of the Vela SNR, \citealt{tml93}) radius of the remnant strongly depends on the choice of the ambient density value, $n_{ISM}$. We set $n_{ISM}=0.5$ cm$^{-3}$, because with this value (and with the chosen values of $K$, $M_{ej}$, $n_{ISM}$, $b$, and $m$), the radius of the shell after 11000 yr is $R_{shell}\sim 5.4\times10^{19}$ cm, in good agreement with the observed radius of the Vela SNR, that ranges between $5\times10^{19}$ cm and $5.5\times10^{19}$ cm (by assuming a distance of $250$ pc, in agreement with \citealt{bms99}, \citealt{csd99}). 

Our model solves the time-dependent compressible fluid equations of mass, momentum, and energy conservation. In three cases we ran the same simulation with/without thermal conduction and radiative cooling inside our system, as shown in Table \ref{tab:setups}. As for thermal conduction, we considered both the Spitzer and the saturated regimes, while radiative losses (that can play an important role in the Vela SNR, as shown by \citealt{mro06}) were computed for an optically thin thermal plasma. The model equations are described in \citet{mro06} (equations 1-5 therein) and were solved by using the FLASH code (\citealt{for00}).

The computational domain extends over $8\times10^{19}$ cm in the $r$ and $z$ directions.  We use axisymmetric boundary conditions at r = 0, reflection boundary conditions at z = 0, and zero-gradient (outflow) boundary conditions (for v, $\rho$, and $p$) elsewhere. We trace the motion of the ejecta material and of the shrapnel with passive tracers\footnote{Both tracers have zero mass and do not modify the dynamics of the system.}.
Considering the large range in spatial scales of our simulations, we exploited the adaptive mesh capabilities of the FLASH code by adopting up to 10 nested levels of resolution (the resolution increases by a factor of 2 at each level). The refinement$/$derefinement criterion \citep{loh87} follows the gradients of density, temperature, and tracers. The finest spatial resolution is $1.95\times10^{16}$ cm at the beginning of the simulation, therefore there are 230 computational cells per initial radius of the ejecta, and $\sim10$ cells per initial radius of the shrapnel (that varies in the range $1.8-2.9\times10^{17}$ cm). Because of the expansion of the system, the resolution is reduced by a factor of 2 after 2500 yr. We verified that by changing the resolution of our simulations by a factor of 2, the results do not change significantly (see Appendix \ref{convergence} for further details).

\begin{center}
\begin{table}
\begin{center}
\caption{Physical parameters of the model setups (see Sect. \ref{Hydrodynamic modeling}). TC-RL indicates that the simulations were perfomed by including thermal conduction and radiative losses. In all the setups the initial radius of the ejecta is $R^{0}_{ej}=4.5\times10^{18}$ cm.}
\begin{tabular}{@{}lcccc} 
\hline\hline
Model setup     &  $\chi$  &       $R_s$ (cm)     & TC-RL   \\ \hline
$R1/3-CHI10$    &    10    &  $1.5\times10^{18}$  &  No     \\
$R1/3-CHI10-TR$ &    10    &  $1.5\times10^{18}$  &  Yes    \\
$R1/3-CHI20$    &    20    &  $1.5\times10^{18}$  &  No     \\
$R1/3-CHI20-TR$ &    20    &  $1.5\times10^{18}$  &  Yes    \\
$R1/3-CHI50$    &    50    &  $1.5\times10^{18}$  &  No     \\
$R1/6-CHI50$    &    50    &  $0.75\times10^{18}$  &  No     \\
$R1/6-CHI50-TR$ &    50    &  $0.75\times10^{18}$  &  Yes    \\
\hline
\label{tab:setups}
\end{tabular}
\end{center}
\end{table}
\end{center}

\section{Results}
\label{Results}

\subsection{Evolution of the system}
\label{evol}

\begin{figure}
 \centerline{\hbox{          
     \psfig{figure=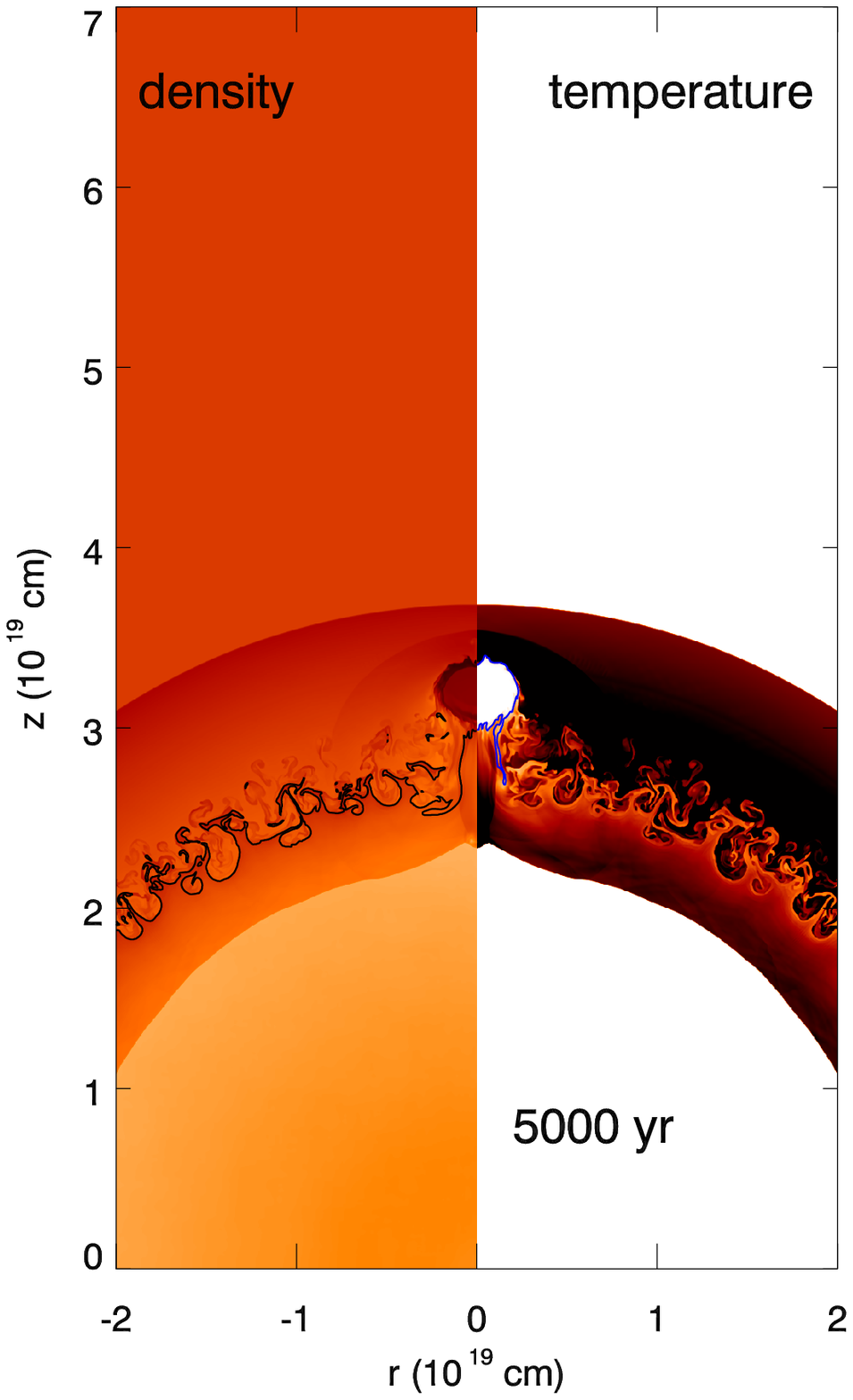,width=3.6cm}     
     \psfig{figure=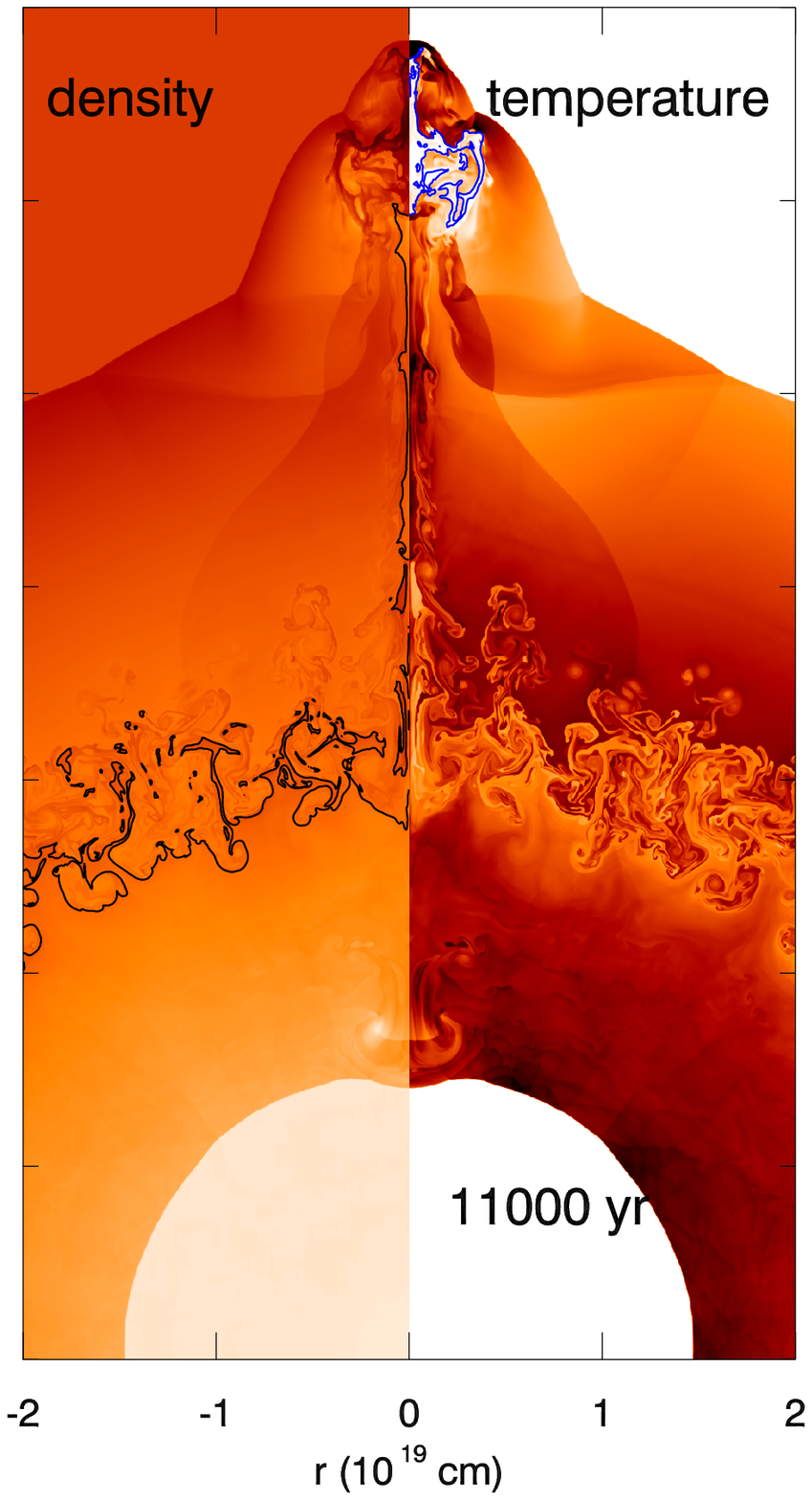,width=3.6cm}     
     }}    
 \centerline{}      
 \centerline{}      
  \centerline{\hbox{	 
      \psfig{figure=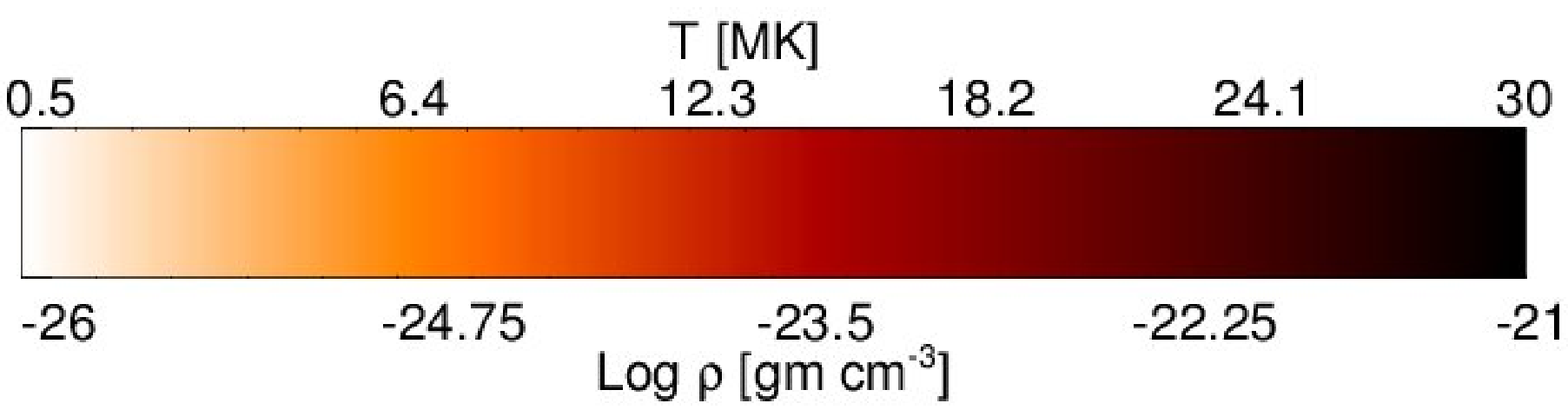,width=5cm}}}
      \caption{\emph{Left panel:} density (\emph{left}) and temperature (\emph{right}) 2-D cross-sections through the $(r,~z)$ plane showing simulation $R1/3-CHI20$ at $t=5000$ yr. The black$/$blue contours enclose the computational cells consisting of the original ejecta$/$shrapnel material by more than $90\%$. The color bar indicates the logaritmic density scale and the linear temperature scale. \emph{Right panel:} same as left panel, for $t=11000$ yr. }
\label{fig:evolhd}
\end{figure}

We first focus on simulation $R1/3-CHI20$.
Figure \ref{fig:evolhd} shows the 2-D cross-sections through the $(r,~z)$ plane of temperature and density at different evolutionary stages of the $R1/3-CHI20$ simulation. The left panel shows the system $5000$ yr after the beginning of the simulation\footnote{In the following, all the ages will be reckoned from the beginning of our simulations. We remind the reader that our initial condition corresponds to $\sim240$ yr after the SN explosion.}, when the shrapnel interacts with the inter-shock region. Rayleigh-Taylor and Richtmyer-Meshkov instabilities are visible as finger-like structures both in the density and temperature maps. At this stage, the knot is partially eroded by the hydrodynamic instabilities and evolves toward a core-plume structure. The core of the knot, however, is still significantly overdense with respect to the surrounding shocked ejecta. The right panel of Fig. \ref{fig:evolhd} shows the shrapnel at $t=11000$ yr, with its characteristic supersonic bow shock protruding beyond the SNR 
main 
shock.


 \begin{figure}
  \centerline{\hbox{	 
      \psfig{figure=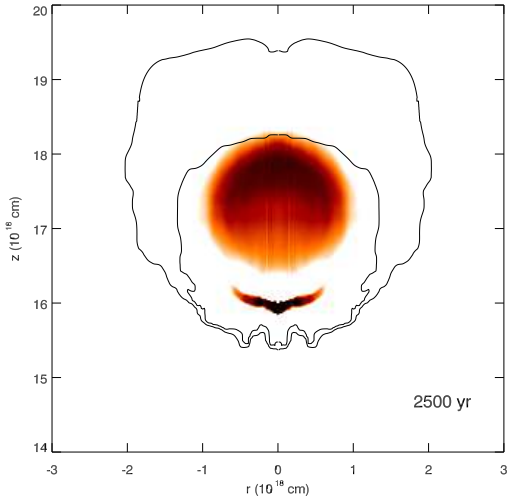,width=6cm}}}
 \centerline{}      
 \centerline{}      
 \centerline{}      
      \caption{Density 2-D cross-sections through the $(r,~z)$ plane for model $R1/3-CHI20$ at $t=2500$ yr. The color scale increases linearly between $10^{-22}$ g$/$cm$^{3}$ and $2\times10^{-22}$ g$/$cm$^{3}$. The contours enclose the computational cells consisting of the original shrapnel material by more than $99\%$ (bold line) and $50\%$ (narrow line).}
\label{fig:knotz}
\end{figure}

 \begin{figure}
  \centerline{\hbox{	 
      \psfig{figure=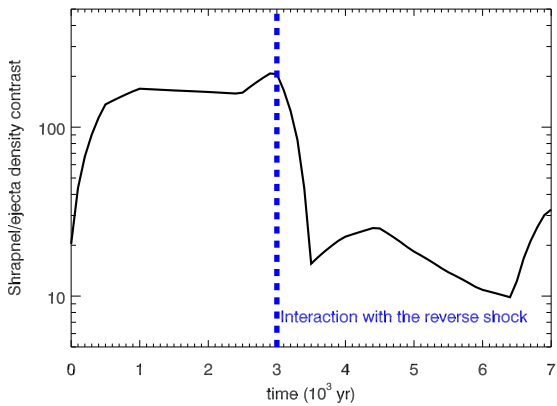,width=\columnwidth}}}
 \caption{Temporal evolution of the shrapnel$/$ejecta density contrast for model $R1/3-CHI20$. The blue horizonthal line marks the beginning of the interaction between the shrapnel and the SNR reverse shock}
\label{fig:chi}
\end{figure}

WC02 found that ejecta knots with density contrast $\chi\le100$ are rapidly fragmented and decelerated in the intershock region and do not even reach the main shock front (these effects being more dramatic for small clumps). Nevertheless, we notice that the value of $\chi$ in WC02 refers to the onset of the interaction between the knot and the reverse shock and that $\chi$ is not constant during the evolution of the system. In the ``free" expansion phase, the density of the shrapnel does not drop down uniformly (as that of the other ejecta does) and the shrapnel undergoes both diffusion and expansion. Figure \ref{fig:knotz} presents a close-up view of the shrapnel density structure at $t=2500$ yr, showing that, while the outer parts of the knot diffuse and mix with the expanding ejecta, its central core remains much denser. The density of the core of the clump drops down much more slowly than that of the spherically expanding ejecta. Therefore, the inhomogeneous rarefaction of the knot makes the density 
contrast between the core of the shrapnel and the expanding ejecta higher, and $\chi$ rapidly increases until the shrapnel reaches the reverse shock. We computed $\chi$ during the expansion phases, by calculating the shrapnel density, $\overline{\rho_s}$, as the average of the density in all the computational cells where the shrapnel content is $>90\%$\footnote{These cells are closer to the center of the knot, and are less affected by the diffusion.}. We then divided $\overline{\rho_s}$ by the ejecta density (along the $r$ axis) at the same distance from the origin as the shrapnel center. Figure \ref{fig:chi} shows the evolution of $\chi$ as a function of time for the $R1/3-CHI20$ simulation.  The figure shows that the ejecta knot reaches $\chi> 100$ as it approaches the reverse shock, hence our results are in agreement with those of WC02. Our model shows that a knot that was only 20 times denser than the surrounding ejecta (at the beginnig of the simulations) can reach the SNR main shock without being 
fragmented in the intershock region and can produce protrusions that are similar to those actually observed in the Vela SNR.

\subsection{Effects of thermal conduction and radiative cooling}
\begin{figure}
 \centerline{\hbox{          
     \psfig{figure=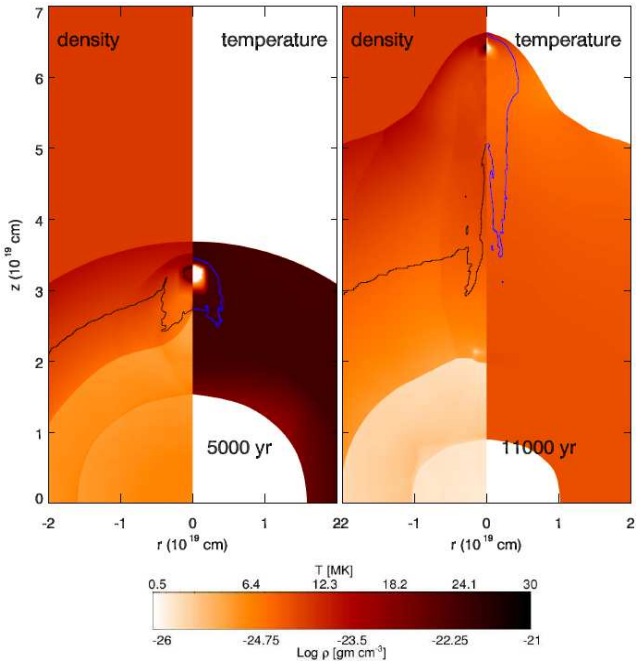,width=\columnwidth} 
     }}
      \caption{Same as Fig. \ref{fig:evolhd} for model $R1/3-CHI20-TR$.}
\label{fig:evol}
\end{figure}

Figure \ref{fig:evol} shows the 2-D cross-sections through the $(r,~z)$ plane of temperature and density at $t=5000$ yr and $t=11000$ yr for the $R1/3-CHI20-TR$ simulation (same parameters as $R1/3-CHI20$, but including radiative cooling and thermal conduction). 
The diffusive thermal conduction completely suppresses the formation and the development of hydrodynamic instabilities and smoothes the temperature and density profiles. This result is in agreement with expectations, as shown below. The characteristic amplitude growth rate, $da/dt$ of a single-mode perturbation of Richtmyer-Meshkov instabilities can be calculated as (see \citealt{ryc60})
\begin{equation}
\frac{da}{dt} = k \Delta v a \overline{A} 
\end{equation}
where $k$ is the perturbation wavenumber, $\Delta v$ is the velocity jump at the instability and $\overline{A}=(\rho_1 - \rho_2)/(\rho_1 + \rho_2)$ is the Atwood number. The characteristic time-scale, $\tau_{inst}=a/(da/dt)$, for the growth of the perturbation is therefore 
\begin{equation}
\tau_{inst} \approx \frac{l}{2\pi \Delta v \overline{A}}\hspace{0.5cm}[{\rm s}]
\end{equation}
where $l$ is the structure size. 
As for the thermal conduction (see \citealt{spi62}),
\begin{equation}
\left(\frac{dE}{dt}\right)_{cond}=\nabla \cdot [\kappa(T)\nabla T]\sim \frac{2}{7}\kappa(T)\frac{T}{l^{2}}
\end{equation}
where $\kappa(T)=5.6\times10^{-7}T^{5/2}$ erg s$^{-1}$ K$^{-1}$ cm$^{-1}$ is the Spitzer's coefficient and $l$ is the characteristic length of temperature variation. therefore, the thermal conduction time-scale is:
\begin{equation}
\tau_{cond}=\frac{7nk}{2(\gamma -1)}\frac{l^{2}}{\kappa(T)}\sim 2.6\times 10^{-9}\frac{nl^{2}}{T^{5/2}}\hspace{0.5cm}[{\rm s}]
\end{equation}

For a characteristic structure with size $l\sim4\times10^{18}$ cm, particle density $n=0.8$ cm$^{-3}$, Atwood number $\overline{A}\sim0.44$, $\Delta v\sim 2\times 10^7$ cm$/$s, $T\sim1.3\times10^7$ K (similar to that shown in Fig. \ref{fig:evolhd}), $\tau_{inst}\sim 1.4\tau_{cond}\sim 2000$ yr. Therefore, the thermal conduction diffusive processes develop faster than the hydrodynamic instabilities and density and temperature inhomogenieties are smoothed out before they can grow.

\begin{figure}
  \centerline{\hbox{	 
      \psfig{figure=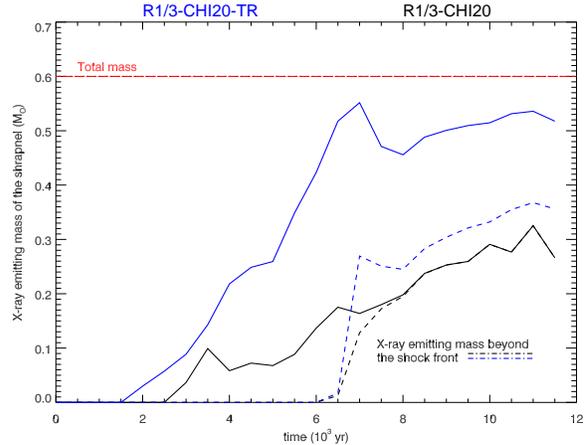,width=\columnwidth,angle=90}}}
 \caption{Temporal evolution of $MX_{shra}$ (mass of the plasma in the computational cells consisting of the original shrapnel material by more than $90\%$ and having a temperature higher than $10^6$ K) for the simulations $R1/3-CHI20-TR$ (blue curves) and $R1/3-CHI20$ (black curves). The dashed lines indicate the part of $MX_{shra}$ that protrudes beyond the SNR shock front.}
\label{fig:shraX}
\end{figure}

The evolution of the position of the shrapnel head and the protrusion it produces to the remnant shock front are similar to those obtained without including thermal conduction and radiative cooling. Nevertheless, the shrapnel evolution is remarkably different from that obtained in the pure HD simulations. In particular, as shown by the blue contours in Fig. \ref{fig:evol}, the ejecta knot is elongated along its direction of motion and rapidly assumes a cometary shape, characterized by a prominent tail which is rich in shrapnel material. After $t=11000$ yr, shrapnel material is present at $\sim10$ pc away from the shrapnel head. 
Moreover, the shrapnel material is efficiently heated by thermal conduction with the surrounding shocked ejecta. Let $MX_{shra}$ be the mass of the plasma in the computational cells consisting of the original shrapnel material by more than $90\%$ and having a temperature higher than $10^6$ K (and therefore emitting thermal X-rays). Figure \ref{fig:shraX} shows the evolution of $MX_{shra}$ as a function of time for the simulations $R1/3-CHI20-TR$ and $R1/3-CHI20$. When thermal conduction is at work, $\sim90\%$ of the original shrapnel mass is heated up to X-ray emitting temperature at $t=11000$ yr (i.~e., at the age of the Vela SNR), while if thermal conduction in inhibited, only $50\%$ of the original mass has temperature higher than $10^6$ K. Figure \ref{fig:shraX} also shows the amount of X-ray mass beyond the shock front. In the pure HD simulation the hot shrapnel material is all beyond the SNR shock front. In the $R1/3-CHI20-TR$ simulation, only part of the X-ray emitting shrapnel is beyond the shock 
front 
and there is a significant fraction 
of 
the ejecta knot material (the shrapnel tail) that is inside the SNR shell and that is expected to emit thermal X-rays (see Sect. \ref{Conclusions}).

We point out that in SNRs the efficiency of thermal conduction can be significantly reduced by the presence of the magnetic field (which is not taken into account in our model). If we assume an organized ambient magnetic field, the thermal conduction is anisotropic, because the conductive coefficient in the direction perpendicular to the field lines is several orders of magnitude lower than that parallel to the field lines, which coincides with the Spitzer's coefficient $\kappa(T)$. The effects of the magnetic-field-oriented thermal conduction in the interaction between shocks and dense clump have been investigated in detail in \citet{obr08}. Because of the high beta of the plasma, the magnetic field lines are expected to envelope the hydrodynamic fingers thus hampering the thermal conduction with the surrounding material. 
At the same time, the magnetic field is expected to be trapped at the top of the ejecta clumps, and this yields to an increase of the magnetic pressure and field tension which limits the growth of hydrodynamic instabilities (see O12 and \citealt{snm12}). 
The basic physics of the interaction between the ejecta knots and the SNR shocks is similar to that for the interaction of planar  shocks with an interstellar cloud (e.~g. WC02) and it has been shown that, in this case, simulations including thermal conduction in an unmagnetized plasma and pure HD simulations are limiting cases that encompass the results obtained with different configurations of the magnetic field (\citealt{obr08}).  
We can then conclude that our simulations provide the two extreme cases that bracket all the possible intermediate scenarios.

\subsection{Effects of the initial conditions}
\label{init}

We study the effects of the initial conditions on the shrapnel evolution with different simulations, as shown in Table \ref{tab:setups}. In particular, we explored two different initial positions of the shrapnel in the ejecta profile ($R_s=1/6~R^{0}_{ej},~1/3~R^{0}_{ej}$) and three different density contrasts ($\chi=10,~20,~50$). 

\begin{figure}
 \centerline{\hbox{          
     \psfig{figure=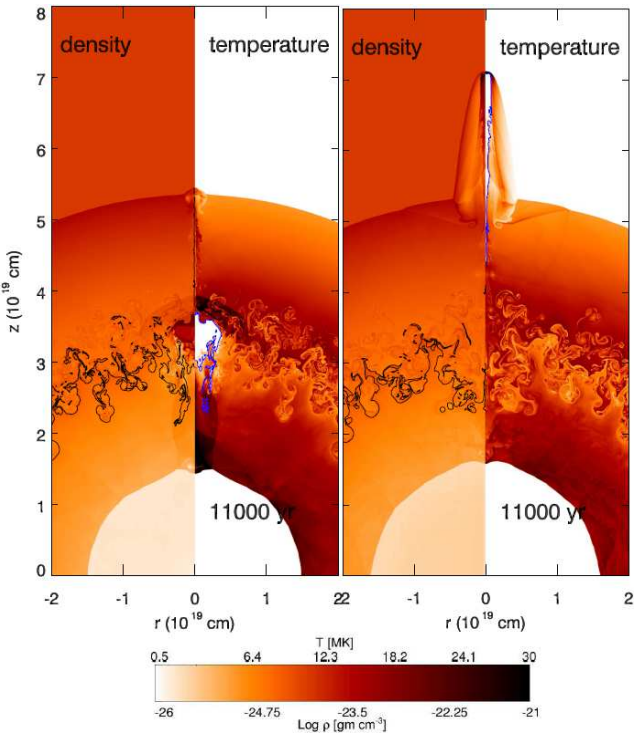,width=\columnwidth}}}     
      \caption{\emph{Left panel:} density (\emph{left}) and temperature (\emph{right}) 2-D cross-sections through the $(r,~z)$ plane showing simulation $R1/6-CHI50$ at $t=11000$ yr. The black$/$blue contours enclose the computational cells consisting of the original ejecta$/$shrapnel material by more than $90\%$. The color bar indicates the logaritmic density scale and the linear temperature scale. \emph{Right panel:} same as left panel, for simulation $R1/3-CHI50$.}
\label{fig:chi50}
\end{figure}

\begin{figure}
  \centerline{\hbox{	 
      \psfig{figure=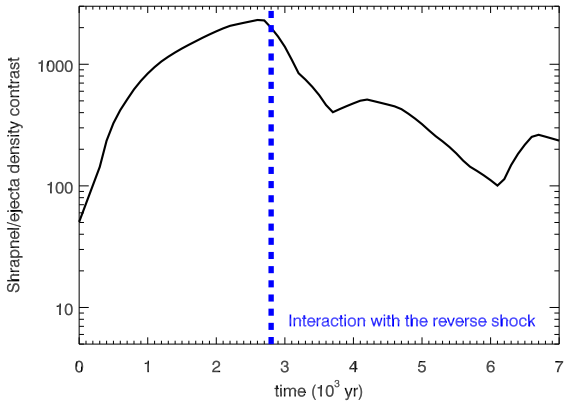,width=\columnwidth}}}
 \caption{Same as Fig. \ref{fig:chi} for model $R1/3-CHI50$.}
\label{fig:chivst2}
\end{figure}

In agreement with WC02, we found that shrapnel formed in the inner ejecta layers (i.~e., those that reach the reverse shock later) produce smaller protrusions. 
In particular, an ejecta knot originating at $R_s=1/6~R^{0}_{ej}$, does not even reach the SNR shock in the time spanned by our simulations. Therefore, our model indicates that shrapnel A-F (all protruding well beyond the Vela main shock) originated in more external layers. Left panel of Fig. \ref{fig:chi50} shows the 2-D cross-sections through the $(r,~z)$ plane of temperature and density at the age of the Vela SNR for the $R1/6-CHI50$ simulation. In this case, the knot is well within the intershock region, even though its initial density contrast ($\chi=50$), was higher than that of the $R1/3-CHI20$ run. However, our models of knots originating at $R_s=1/3~R^{0}_{ej}$ clearly show that denser shrapnel produce deeper protrusions and are more stable against the fragmentation and the deceleration induced by the hydrodynamic instabilities in the inter-shock region (as in WC02)\footnote{In our runs, higher values of $\chi$ correspond to smaller values of the shrapnel cross-section (given that the 
shrapnel mass is fixed to $1.19\times10^{33}$ g in all our simulations), and this concurs in making the knot a more penetrating bullet.}.
Fig. \ref{fig:chi50}, right panel, shows the 2-D cross-sections through the $(r,~z)$ plane of temperature and density at the age of the Vela SNR for the $R1/3-CHI50$ simulation. As expected, the shrapnel head is much further away from the SNR shell than in the $R1/3-CHI20$ case (shown in the right panel of Fig. \ref{fig:evolhd}). We computed the time evoulution of $\chi$ for the $R1/3-CHI50$ run by following the procedure described in Sect. \ref{evol}. We found that, in this case, the ejecta knot reaches the reverse shock with a very high density contrast ($\chi>1000$, see Fig. \ref{fig:chivst2}), thus producing a very prominent protrusion in the SNR.


\begin{figure}
  \centerline{\hbox{	 
      \psfig{figure=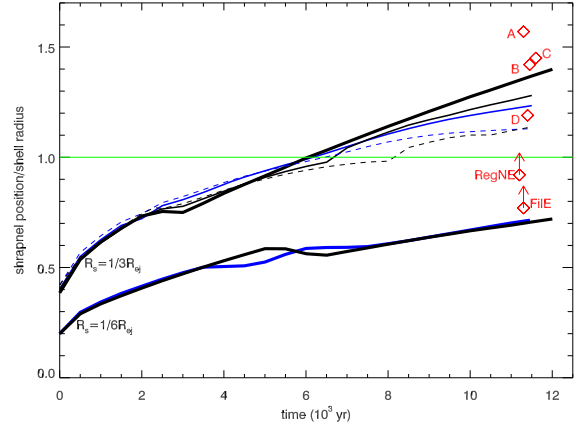,width=\columnwidth,angle=90}}}
 \caption{Position of the shrapnel head (in units of the shell radius $R_{shell}$) as a function of time for all the simulations listed in Table \ref{tab:setups}. Black curves indicate pure hydrodynamic models, while blue curves indicate simulations including thermal conduction and radiative cooling. Dashed/solid/thick lines indicate $\chi=10/30/50$, respectively. The red diamonds show the projected positions of Shrapnel A-D (\citealt{aet95}) and of the ejecta knots FilE and RegNE (\citealt{mbr08}).}
\label{fig:shrapos}
\end{figure}

Figure \ref{fig:shrapos} shows the position of the shrapnel head (in units of the shell radius) as a function of time for all our simulations. Figure \ref{fig:shrapos} also shows the projected positions of Shrapnel A-D (\citealt{aet95}) and of the ejecta knots FilE and RegNE (\citealt{mbr08}) with respect to the position of the shock front in the Vela SNR. These values were calculated by approximating the Vela SNR as a circular shell with angular radius $211'$ and center with coordinates $\alpha_{J2000}=8^h36^m19.8^s,~\delta_{J2000}=-45^\circ24'45''$.  Models including the effects of radiative cooling and thermal conduction (blue curves in Fig. \ref{fig:shrapos}) do not provide significant differences with respect to pure HD models (black curves) in terms of the shrapnel position.

By considering all the simulations with $R_s=1/3~R^{0}_{ej}$, we find that the position of the head of the knot, $R_h$, after $\sim11000$ yr ranges between $\sim1.1~R_{shell}$ (for $\chi=10$) and $\sim1.4~R_{shell}$ (for $\chi=50$). These values are similar to those observed for Shrapnel B, C, and D.

Shrapnel A, E, and F (Shrapnel E, F are not shown in  Fig. \ref{fig:shrapos}) instead, have $R_h=1.57~R_{shell}$, $R_h=1.88~R_{shell}$, and $R_h=1.91~R_{shell}$, respectively. These values are much larger than those obtained in our simulations. Our results clearly suggest that these large distances from the shell can be produced by ejecta knots originating in outer ejecta layers. 
An alternative possibility is that the original density contrast for these shrapnel was $>50$. Nevertheless, Fig. \ref{fig:shrapos} shows that the final position of an ejecta clump is more sensitive to its initial position and depends only weakly on $\chi$. In fact, by varying the initial position of the knot by a factor of two, we found that its final position varies approximately by the same factor, while a variation of the initial density contrast by a factor of five, only determines a 30\% variation in the final position. Therefore it is more likely that Shrapnel A, E, and F were produced at $R_s>1/3~R^{0}_{ej}$.

The two simulations with $R_s=1/6~R^{0}_{ej}$ show that the ejecta knots originating in the inner ejecta layers do not reach the forward shock, even for the highest density contrast ($\chi=50$). The position of the head of the knot predicted by our simulations is in agreement with that observed for FilE (\citealt{mbr08}). X-ray emitting ejecta knots have been observed, in projection, inside the Vela SNR shell (e.~g., RegNE and FilE, \citealt{mbr08}). However, we point out that the actual position of these ``internal'' shrapnel can be well outside the SNR shell, and therefore the values reported in Fig. \ref{fig:shrapos} might be considered as lower limits.

\section{Discussions and conclusions}
\label{Conclusions}

The structure of the ejecta in a SNR contains the imprint of the metal-rich layers inside the progenitor star, and may help to understand the processes occurring in the latest stage of stellar evolution. We performed a set of hydrodynamic simulations to study the evolution of the ejecta knots in the Vela SNR. 

\begin{figure}
  \centerline{\hbox{	 
      \psfig{figure=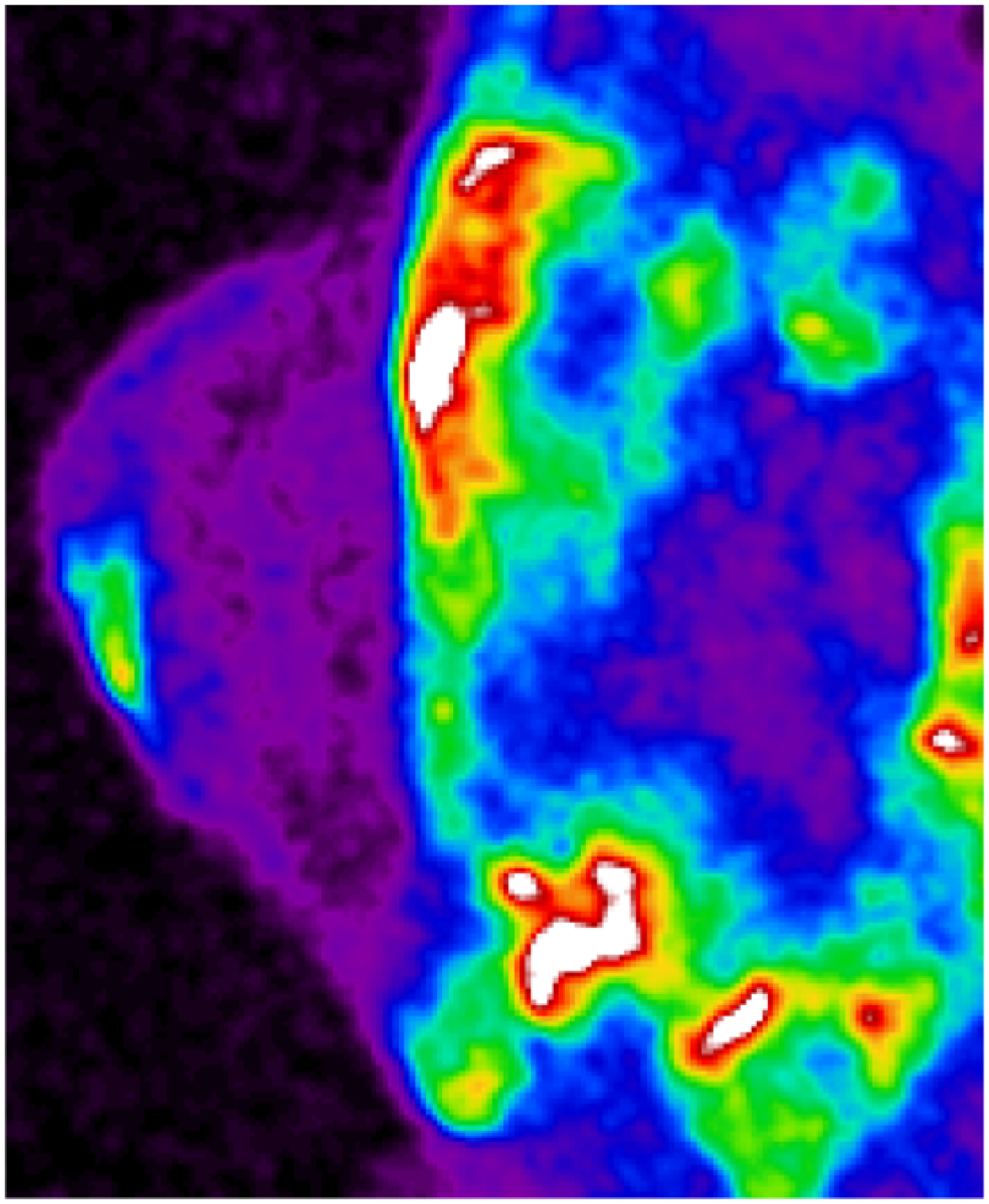,width=\columnwidth}}}
  \centerline{\hbox{	       
      \psfig{figure=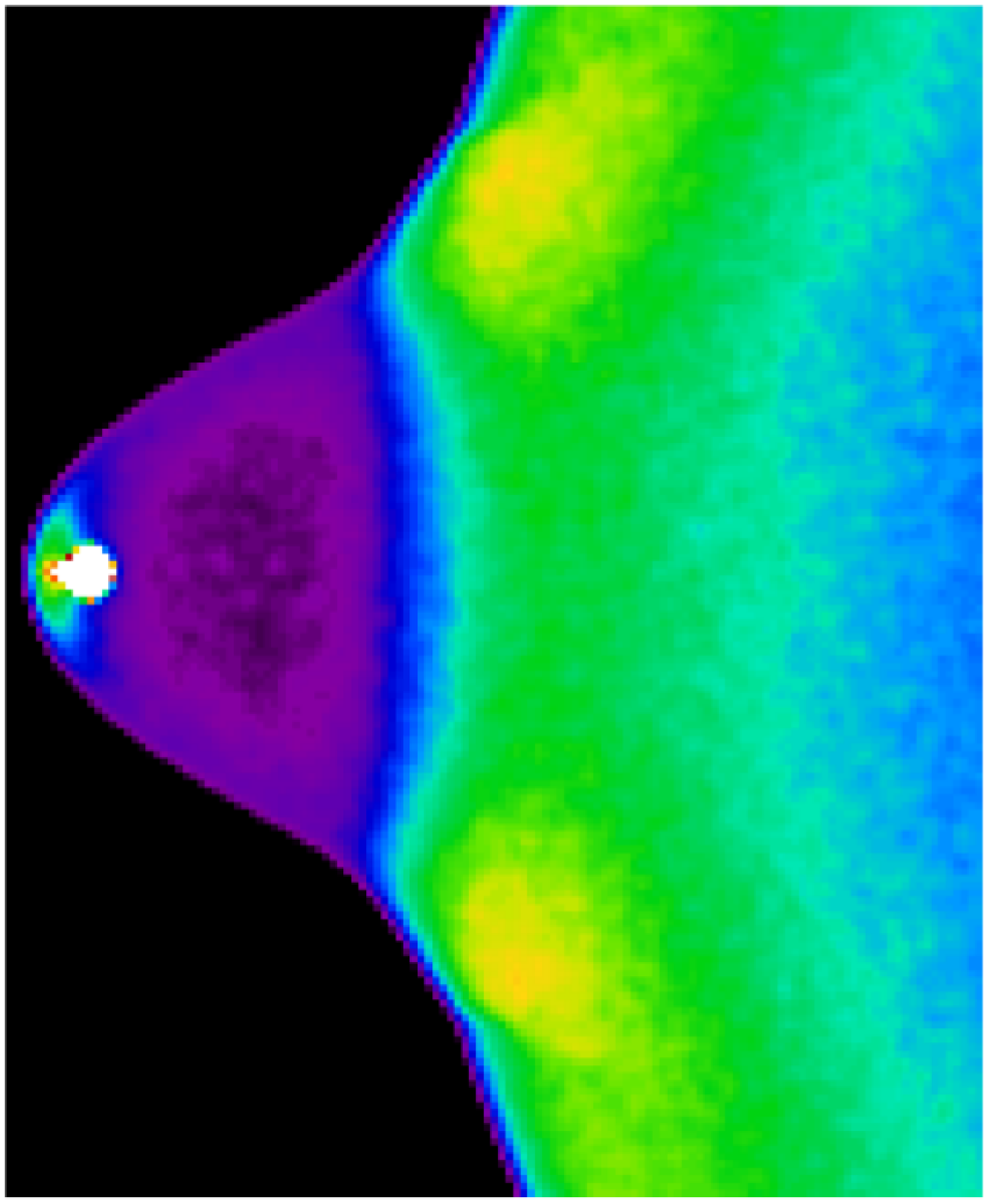,width=\columnwidth}}}
 \caption{\emph{Upper panel:} Rosat All-Sky Survey image of Vela Shrapnel D in the $0.1-2.4$ keV energy band. The bin-size is $1.5'$ and the image has been smoothed through a convolution with a Gaussian with $\sigma=3$ pixels. North is up and East is to the left. \emph{Lower panel:} Synthesized X-ray emission in the $0.1-2.4$ keV band for the simulation $R1/3-CHI20-TR$ at $t=11000$ yr (see Sect. \ref{Conclusions}).}
\label{fig:shraD}
\end{figure}

We found that moderately overdense clumps (initial density contrast $\chi\sim10$) can produce protrusions in the SNR shell similar to those observed for the Vela shrapnel. 
WC02 found that only clumps that reach the reverse shock with density contrast $\chi\ge100$ can reach the main shock front and produce significant protrusions. This criterium is fulfilled in all our simulations. In fact, the (initially) moderately overdense clump experiences diffusion in the ``free" expansion phase, and, while its outer parts mix with the surrounding ejecta, its central core remains much denser. This inhomogeneous rarefaction makes the density contrast between the core of the shrapnel and the expanding ejecta higher as the remnant evolves, and $\chi$ reaches values $\sim100-1000$ at the interaction with the reverse shock.

In particular, a knot with initial $\chi=20$ and $R_s=1/3~R^{0}_{ej}$ (simulations $R1/3-CHI20$ and $R1/3-CHI20-TR$) can explain the observed features associated with the Vela Shrapnel D. 
Figure \ref{fig:shraD} shows the the $ROSAT$ All-Sky Survey image of the Vela Shrapnel D in the $0.1-2.4$ keV energy band, compared with a synthesis of the X-ray emission in the $0.1-2.4$ keV band derived from the $R1/3-CHI20-TR$ simulation at $11000$ yr. The synthesized X-ray map has been obtained as in \citet{mro06}: we produced the 3D map of the emission measure and temperature in the cartesian space ($x',~y',~z'$), where the $y'$ axis corresponds to the direction of the line of sight and is perpendicular to the $(r,~z)$ plane. We derived the distribution $EM(x',~z')$ vs. $T(x',~z')$ by considering all the contrbutions along the line of sight, for each $(x',~z')$. We then synthesized the map of the X-ray emission using the MEKAL spectral code (\citealt{mgv85}, \citealt{mlv86}, \citealt{log95}), assuming a distance of $250$ pc, and an interstellar column density $N_{H}=2\times 10^{20}$ cm$^{-2}$. Finally, we degraded the spatial resolution of the synthesized X-ray map to match the resolution of the ROSAT 
image and randomized the map by assuming Poisson statistics for the counts in each image bin.

Figure \ref{fig:shraD} shows that the bright X-ray spot at the apex of the shrapnel bow-shock and the enhanced X-ray luminosity at the base of the protrusion can be explained by our model as local enhancements of the plasma emission measure. The overall shape of the observed features is also reproducted by our model. Moreover, the mass of the X-ray emitting shrapnel predicted by our model is in agreement with that measured for Shrapnel D. In fact, it has been estimated that the X-ray emitting mass (considering only the bulge above the Vela SNR bow shock) is $\sim0.1M_\odot$ \citep{kt05} i.~e., the same order of magnitude as that obtained in our simulation (see dashed curves in Fig. \ref{fig:shraX}).
Vela Shrapnel D is therefore consistent with being originated by an ejecta knot 20 times denser than the surrounding ejecta and initially located at $1/3~R^{0}_{ej}$. As shown in Fig. \ref{fig:shrapos}, we found a degeneracy in the space of the initial parameters, and similar results can be obtained also by enhancing/decreasing the density contrast and decreasing/increasing the initial distance of the knot to the explosion center. However, as explained in Section \ref{init}, the final position of the shrapnel is much more sentitive to its initial position in the ejecta profile. In conclusion, according to our model, the original position of Shrapnel D should not differ much from $1/3~R^{0}_{ej}$.

As shown in Sect. \ref{init}, our model suggests that Shrapnel B, C, and D were all originated at $\sim1/3~R^{0}_{ej}$. This conclusion is in agreement with the results of X-ray data analysis that show that Shrapnel B and D have similar abundance patterns (\citealt{kt05}, \citealt{yk09}). The fragment RegNE, which appears inside the Vela shell, has similar abundances as Shrapnel D \citep{mbr08} and its position is compatible with an origin in the same layer where Shrapnel B, C, and D were generated\footnote{In this case, its projected distance from the center of the SNR  must be much smaller than its actual distance.}.

In Sect. \ref{init}, we also pointed out that it is highly unlikely that Shrapnel A originated in the same ejecta layer as Shrapnel D. Indeed, the abundance patterns observed in Shrapnel A are remarkably different from that observed in Shrapnel B and D (\citealt{kt06}), thus suggesting a different location of this knot in the ejecta profile. Nevertheless, the Si:O ratio is much higher in Shrapnel A than in Shrapnel D and this indicates that Shrapnel A comes from a region of the progenitor star below that of the Shrapnel D (e.~g. \citealt{tk06}), at odds with our predictions. 
As explained above, an unrealistically high initial $\chi$ is required for an inner shrapnel to overcome outer knots, if we assume that the initial velocity profile of the ejecta increases linearly with their distance from the center. A possible solution for this puzzling result is that the Si burning layer (or part of it) has been ejected with a higher initial velocity, e.~g., as a collimated jet. 
It is noteworthy to remark that in other core-collapse SNRs the Si-rich ejecta may show a very peculiar jet-counterjet structure. The  well known case of Cas A has been studied in detail thanks to a very long Chandra observation (\citealt{hlb042}) showing a jet (with a  weaker counterjet structure) composed mainly of Si-rich plasma. \citet{lhr06}  have performed X-ray spectral analysis of several  knots in the jet and concluded that the origin of this  interesting morphology is due to an explosive jet and it is not arising because of an interaction with a cavity or other ISM/CSM peculiar structure. Therefore a jet origin for the Si-rich knots is sound.

Finally, we investigated the effects of thermal conduction, finding that it determines an efficient ``evaporation" of the ejecta knot and accelerate its mixing with the surrounding medium. Moreover, it affects the shrapnel morphology, producing the formation of a long, metal-rich tail. Enhanced metallicities have been observed in the tails of Shrapnel A, B, and D and the abundance analysis performed on the X-ray spectra clearly suggests an efficient mixing of the ejecta knots with the surrounding medium (\citealt{kt05, kt06}, and \citealt{yk09}). These results are in qualitative agreement with our findings. A quantitative comparison between models and observations requires a forward modeling approach, consisting in the synthesis of the X-ray spectra from the simulations and a detailed comparison between the synthesized observables and the corresponding observations (e.~g., through a spatially resolved spectral analysis, as in \citealt{mro06}, e. g.). Moreover it will also be important for future 
models to include some seed magnetic fields in the simulations to study how tangled the field becomes and what this implies for thermal conduction. These further studies are beyond the scope of this paper.

In conclusion, our hydrodynamic modelling of ejecta knots in the Vela SNR allowed us to find that: i) the observed shrapnel can be the results of moderate density inhomogeneities in the early ejecta profile; ii) the evolution of a shrapnel in the SNR is very sensitive to its initial position and depends much less (but does depend) on the initial density contrast; iii) thermal conduction plays an important role and explains the efficient mixing of the ejecta knots observed in X-rays.

\section*{Acknowledgments}
We thank the anonymous referee for their constructive suggestions to improve the paper. 
The software used in this work was in part developed by the DOE-supported ASC / Alliance Center for Astrophysical Thermonuclear Flashes at the University of Chicago. The simulations discussed in this paper have been performed on the HPC facility at CINECA, Italy, and on the GRID infrastructure of the COMETA Consortium, Italy. 

\bibliographystyle{mn2e}

\appendix
\section{Test on spatial resolution}
\label{convergence}

\begin{figure}
  \centerline{\hbox{	 
      \psfig{figure=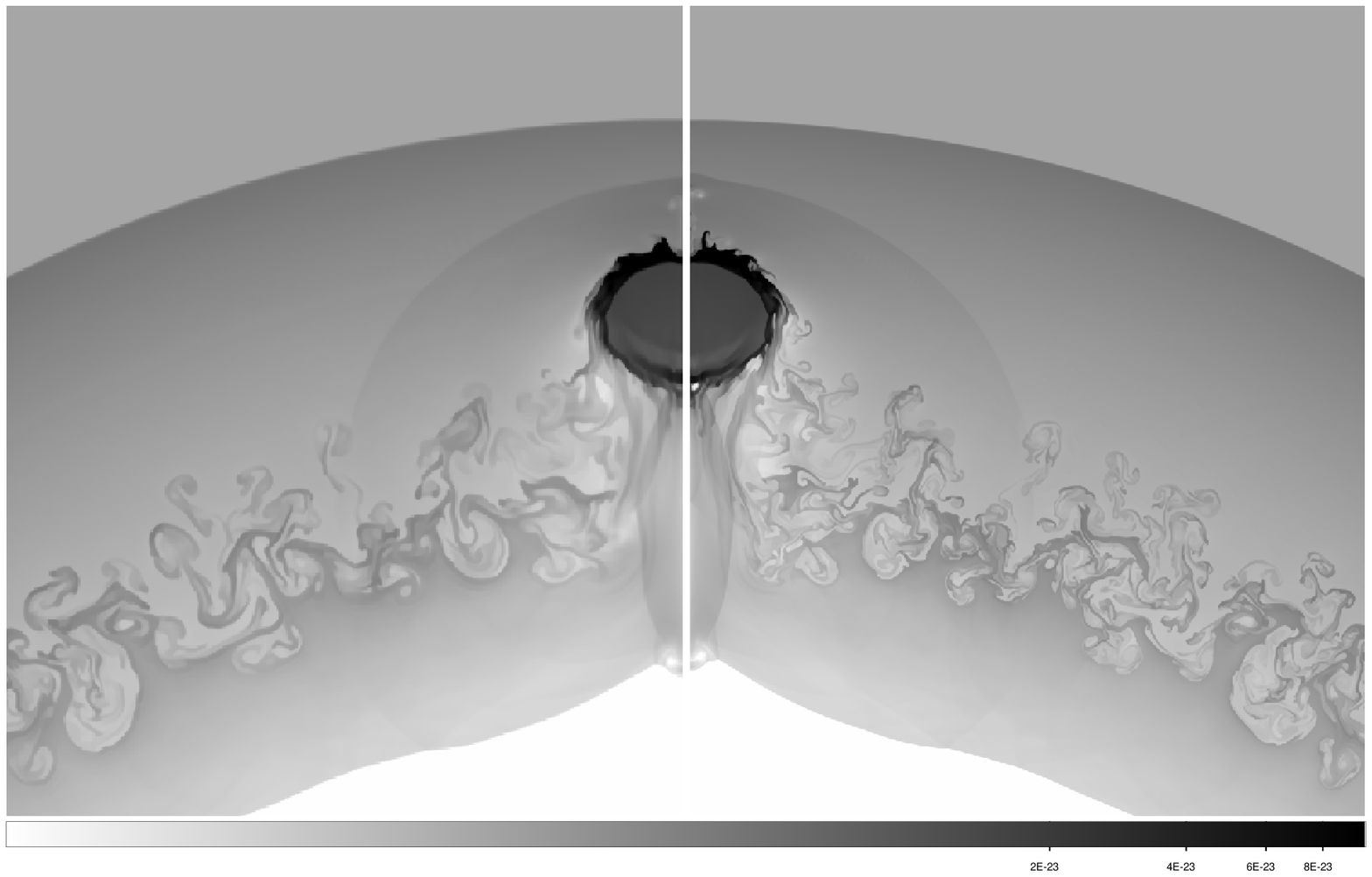,width=\columnwidth}}}
\caption{Density 2-D cross-sections through the $(r,~z)$ plane for simulation $R1/3-CHI20$ (\emph{left panel}) and $R1/3-CHI20-hiRES$ (\emph{right panel}) at $t=5000$ yr. The color bar indicates the logaritmic density scale  and ranges between $10^{-25}$ g cm$^{-3}$ and $10^{-22}$ g cm$^{-3}$. }
\label{fig:hilores}
\end{figure}

As explained in Sect. \ref{initcond}, the spatial resolution of our simulations is reduced by a factor of 2 at $t>2500$ yr. To check whether this resolution is sufficient to capture the basic evolution of the system, we repeated simulation $R1/3-CHI20$ by mantaining the finest level of resolution at its initial value ($1.95\times10^{16}$ cm, hereafter simulation $R1/3-CHI20-hiRES$) for the whole run. We verified that simulations $R1/3-CHI20$ and $R1/3-CHI20-hiRES$ yield very similar results and all the quantities discussed in the paper (position of the shrapnel as a function of time, X-ray emitting mass of the shrapnel, etc.) do not change significantly. Figure \ref{fig:hilores} shows the 2-D cross-sections of the density through the $(r,~z)$ plane obtained for simulations $R1/3-CHI20$ and $R1/3-CHI20-hiRES$ at $t=5000$ yr. This evolutionary stage is the most critical, since the system is still relatively small and the ejecta knots interact with the hydrodynamic instabilities in the intershock region. Figure 
\ref{fig:hilores} clearly prove that simulation $R1/3-CHI20$ already provides a very accurate description of the system and run $R1/3-CHI20-hiRES$ does not introduce any major differences. 

\end{document}